\documentclass[10pt]{sigplanconf}

\usepackage{url}               
\usepackage{listings}          
\usepackage{algorithm}
\usepackage{algpseudocode}
\usepackage{algorithmicx}
\usepackage[usenames,dvipsnames,svgnames,table]{xcolor}
\usepackage{hyperref}
\usepackage{amssymb}
\usepackage{graphicx}

\newcommand{\TODO}[1]{[\textsl{#1}]}
\newcommand{\code}[1]{\lstinline{#1}}
\newcommand{\package}[1]{\lstinline{#1}}

\newcommand{\etal}{\textit{et al.}}

\begin{document}

\lstdefinelanguage{Julia}{
   morekeywords={end,
	begin,  while, if, for, try, return,
        break, continue, function, stagedfunction,
        macro, quote, let, local, global, const,
        abstract, typealias, type, bitstype, immutable,
        ccall, do, module, baremodule, using, import, elseif,
        else, export, importall},
   sensitive=true,
   showspaces=false,
   showstringspaces=false,
   morecomment=[l]\#,
   morecomment=[n]{\#=}{=\#},
   morestring=[b]",
}
\lstset{
   basewidth=0.5em,
   basicstyle=\footnotesize\ttfamily,
   keywordstyle=\color{OliveGreen}\bfseries,
   commentstyle=\color{gray}\itshape,
   breaklines=true,
   language=Julia,
   mathescape=true,
   tabsize=2,
}


\preprintfooter{}   

\title{Fast Flexible Function Dispatch in Julia}
\authorinfo{Jeff Bezanson}
        {Massachusetts Institute of Technology and\\
        Julia Computing}
        {jeff.bezanson@gmail.com}
\authorinfo{Jake Bolewski}
        {Massachusetts Institute of Technology and\\
        TileDB}
        {jake.bolewski@gmail.com}
\authorinfo{Jiahao Chen}
        {Massachusetts Institute of Technology and\\
        Capital One Financial}
        {cjiahao@gmail.com}

\maketitle

\begin{abstract}
  Technical computing is a challenging application area for programming
  languages to address. This is evinced by the unusually large number of
  specialized languages in the area (e.g. MATLAB~\cite{matlab},
  R~\cite{rlang}), and the complexity of common software stacks, often
  involving multiple languages and custom code generators.
  We believe this is ultimately due to key characteristics of the domain:
  highly complex operators, a need for extensive code specialization for
  performance, and a desire for permissive high-level programming styles
  allowing productive experimentation.

  The Julia language attempts to provide a more effective structure
  for this kind of programming by allowing programmers to express complex
  polymorphic behaviors using dynamic multiple dispatch over
  parametric types. The forms of extension and reuse
  permitted by this paradigm have proven valuable for technical computing.
  We report on how this approach has allowed domain experts to express
  useful abstractions while simultaneously providing a natural path to better
  performance for high-level technical code.
\end{abstract}




\section{Introduction}
Programming is becoming a growing part of the work flow of those working in the physical scientists. [Say something comparing the number of type of programmers in some previous decade to now?] These programmers have demonstrated that they often have needs and interests different from what existing languages were designed for.

In this paper, we focus on the phenomenon of how dynamically typed languages such as Python, Matlab, R, and Perl have become popular for scientific programming. Dynamic languages features facilitate writing certain kinds of code, use cases for which occur in various technical computing applications.
Holkner~\etal's analysis of Python programs~\cite{Holkner2009} revealed the use of dynamic
features across most applications, occurring mostly at startup for data I/O, but
also throughout the entire program's lifetime, particularly in programs
requiring user interactivity. Such use cases certainly appear in technical
computing applications such as data visualization and data processing scripts.
Richards~\etal's analysis of JavaScript programs~\cite{Richards2010} noted that dynamic
features of JavaScript were commonly use to extend behaviors of existing types
like arrays. Such use cases are also prevalent in technical computing
applications, to imbue existing types with nonstandard behaviors that are
nonetheless useful for domain areas.

An issue that arises with dynamically typed languages is performance.
Code written in these languages is difficult to execute efficiently~\cite{Joisha2001,Joisha2006,Seljebotn2009}.
While it is possible
to greatly accelerate dynamic languages with various techniques, for
technical computing the problem does not stop there. These systems
crucially depend on large libraries of low-level code that provide array
computing kernels (e.g. matrix multiplication and other linear algebraic
operations). Developing these libraries is a challenge, requiring a range of
techniques including templates, code generation, and manual code
specialization. To achieve performance users end up having to transcribe their prototyped codes into a lower level static language, resulting in duplicated effort and higher maintenance costs. [Talk about how this happens in Cython.]

We have designed the Julia programming language~\cite{Bezanson2012, Bezanson2014} allows the programmer to combine dynamic types with static method dispatch. We identify method dispatch as one of the key bottlenecks in scientific programming. As a solution we present a typing scheme that allows the programmer to optionally provide static annotations for types. In this paper, we describe Julia's multiple dispatch type system, the annotation language, and the data flow algorithm for statically resolving method dispatch. By analyzing popular packages written in Julia, we demonstrate that 1) people take advantage of Julia's type inference algorithm, 2) the Julia compiler can statically resolve X\% of method dispatch in Julia programs, and 3) static dispatch in Julia provides performance gains because it enables inlining. [Also show that statically resolving dispatch provides speedups?]



Our main contributions are as follows:
\begin{itemize}
\item We present the Julia language and its multiple dispatch semantics.
\item We describe the data flow algorithm for resolving Julia types statically.
\item We analyze X Julia packages and show that Y
\item (Talk about how much static dispatch speeds things up?)
\item (Talk about how much of the code actually gets annotated?)
\end{itemize}

\section{Introductory Example}

We introduce Julia's multiple dispatch type system and dispatch through an example involving matrix multiplication. We show how Julia's type system allows the programmer to capture rich properties of matrices and how Julia's dispatch mechanism allows the programmer the flexibility to either use built-in methods or define their own.

\subsection{Expressive Types Support Specific Dispatch}
The Julia base library defines
many specialized matrix types to capture properties
such as triangularity, Hermitianness or bandedness. Many specialized
linear algebra algorithms exist that take advantage of such information.
Furthermore some matrix properties lend themselves to multiple representations.
For example, symmetric matrices may be stored as ordinary matrices, but only
the upper or lower half is ever accessed. Alternatively, they may be stored in
other ways, such as the packed format or rectangular full packed format, which
allow for some faster algorithms, for example, for the Cholesky
factorization~\cite{Gustavson2010}.
Julia supports types such as \lstinline|Symmetric| and \lstinline|SymmetricRFP|, which encode information about matrix properties and their storage format.
In contrast, in the LAPACK (Linear Algebra Package)~\TODO{cite} Fortran library for numerical linear algebra, computations
on these formats are distinguished by whether the second and third letters of
the routine's name are \lstinline|SY|, \lstinline|SP| or \lstinline|SF|
respectively.

Expressive types support specific dispatch.
When users use only symmetric matrices, for example, in code that
works only on adjacency matrices of undirected graphs,
it makes sense to construct \lstinline|Symmetric| matrices explictly.
This allows Julia to dispatch directly to specialized methods for \lstinline|Symmetric|
types. An example of when this is useful is \lstinline|sqrtm|, which computes the
principal square root of a matrix:
\begin{lstlisting}
A = Symmetric([1 0 0; 0 0 1; 0 1 0])
B = sqrtm(A)
\end{lstlisting}
In general, the square root can be computed
via the Schur factorization of a matrix~\cite{Golub2013}. However, the square
root of a \lstinline|Symmetric| matrix can be computed faster and more stably
by diagonalization to find its eigenvalues and
eigenvectors~\cite{Higham2008,Golub2013}. Hence, it is always advantageous to
use the spectral factorization method for \lstinline|Symmetric| matrices.



\subsection{Dynamic Dispatch}
\begin{figure}
	\centering
	\includegraphics[width=\columnwidth]{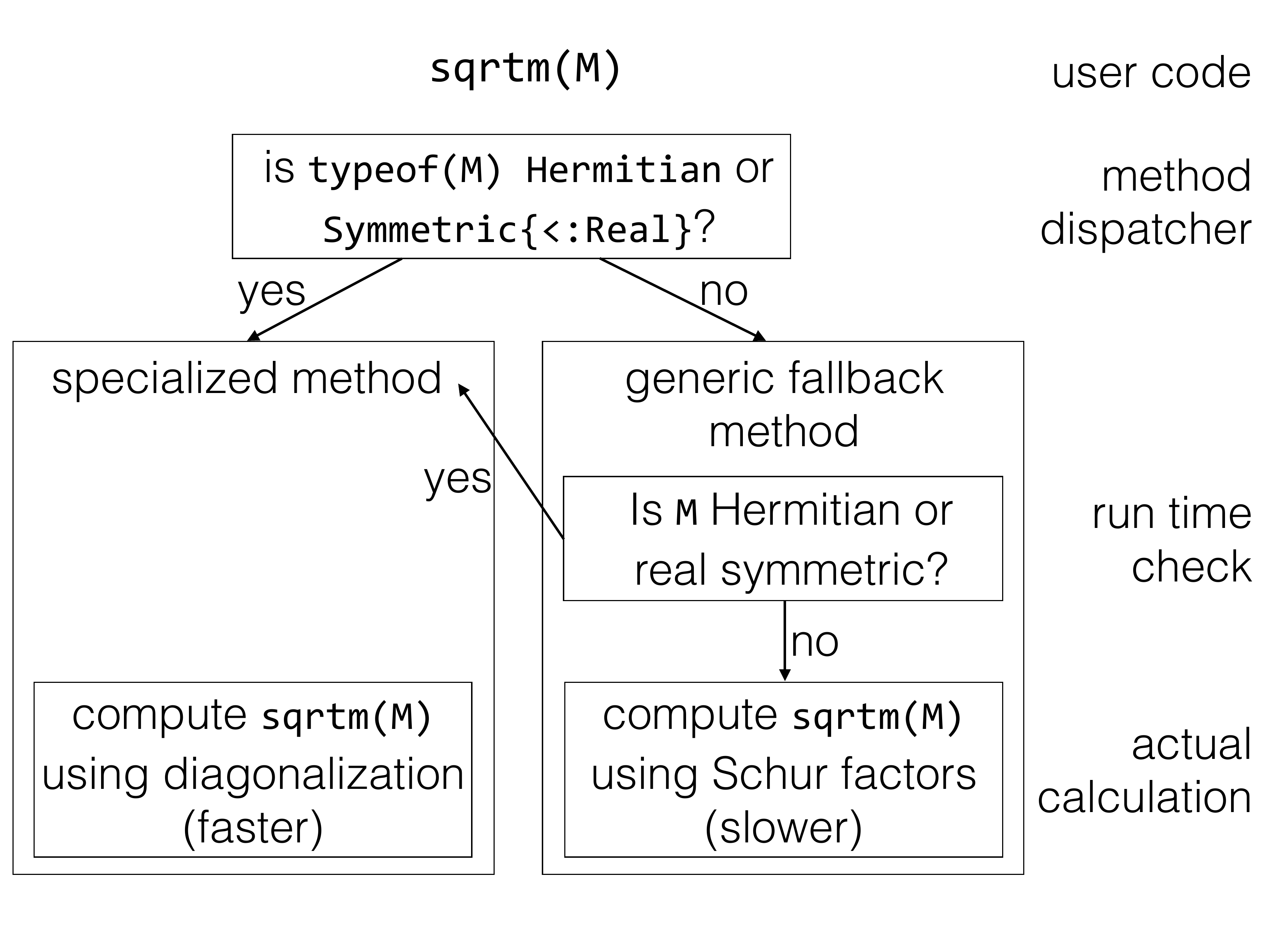}
	\caption{Dynamic dispatch and multimethods for the matrix square root
		function \texttt{sqrtm}, showing that the specialized algorithm
		can be run either from a static decision from the method
		dispatcher based on the input type, or a dynamic decision from
		a run-time check based on the value.}
	\label{fig:sqrtm}
\end{figure}

Julia's implementation of \lstinline|sqrtm| uses the type to check whether to dispatch on the specilized method or to fall back on Schur factorization:\footnote{
The code listing is taken directly from the Julia base library, with minor
changes for clarity.}
\begin{lstlisting}
function sqrtm{T<:Real}(A::StridedMatrix{T})
	#If symmetric, use specialized method
	issym(A) && return sqrtm(Symmetric(A))

	#Otherwise, use general method
	SchurF = schurfact(complex(A))
	R = full(sqrtm(Triangular(SchurF[:T], :U, false)))
	retmat = SchurF[:vectors]*R*SchurF[:vectors]'

	#If result has no imaginary component, return a matrix of real numbers
	all(imag(retmat) .== 0) ? real(retmat) : retmat
end
\end{lstlisting}
We summarize the high-level behavior of \lstinline|sqrtm| in Figure~\ref{fig:sqrtm}.
The general method first checks if the input matrix is symmetric, which is fast
compared to the actual computation of the square root. If the matrix is found to
be symmetric, it is wrapped in a \lstinline|Symmetric| constructor, which allows
Julia to dispatch to the specialized method. Otherwise, the next few lines
compute the square root using the slower method.
Thus a user-level \lstinline|sqrtm(A)| call will in
general be dynamically dispatched, but can ultimately call the same
performant kernel if the argument happens to be symmetric.
The type of result returned by \lstinline|sqrtm| depends
on run-time checks for symmetry and real-valuedness.

\subsection{Resolving Dispatch Statically}
This example illustrates a pattern characteristic of technical
computing environments: dynamic dispatch on top of statically-typed
performance kernels.
However the difference here is that both components can be expressed
in the same language, and with the same notation.
\TODO{Talk about the static dispatch algorithm.}


\TODO{Incorporate Jeff's example from his thesis.}

\subsection{Polymorphic types in Julia}
\TODO{This part of the story seems to not be front and center anymore. What should we do about it?}

Code specialization is only half of the story.
The other half is extensibility --- adding new data types and
method definitions.
Julia's approach to extensibility is designed to take advantage of
specialization, by allowing methods to be defined for combinations
of structured types.

The method shown above demonstrates two kinds of polymorphism supported in
Julia. First, the input \lstinline|A| is annotated (\lstinline|::|) with the
type \lstinline|StridedMatrix|, which is an abstract type, not a concrete type.
A strided matrix is one that has a Fortran-compatible storage format, i.e.\ it
is stored in a contiguous block of memory and whose dimensions can be described
by a dope vector. The subtypes of \lstinline|StridedMatrix| in the base
library are \lstinline|Matrix| (ordinary matrices), and subarray types defining
views into strided matrices, whose referenced elements may not be fully
contiguous in memory. Thus, this \lstinline|sqrtm| method would be
dispatched upon by both \lstinline|Matrix| objects and subarrays, by virtue of
both types being subtypes of \lstinline|StridedMatrix|.

The second kind of polymorphism shown by the signature
\lstinline|sqrtm{T<:Real}(A::Matrix{T})| is parametric polymorphism. This
signature defines a family of related methods, one for each kind of
\lstinline|Matrix| containing a numeric type \lstinline|T| that is a subtype
(\lstinline|<:|) of \lstinline|Real|. This definition encompasses separate
definitions for matrices of type \lstinline|Matrix{Int32}| (matrices of 32-bit
integers), \lstinline|Matrix{Float64}| (matrices of 64-bit floating point real
numbers), and even \lstinline|Matrix{Rational{BigInt}}| (matrices of rational
numbers where the numerators and denominators are arbitrary precision
integers). Matrices containing other types, such as
\lstinline|Matrix{Complex{Float64}}|, would not dispatch on this family of methods.

The two kinds of polymorphism allow families of related methods to be defined
concisely, which allow highly generic code to be written. Additionally, code for
multiple algorithms can coexist within the same generic function, with the actual
choice of which code to run being determined by method dispatch. Furthermore,
the same performant kernel can be called even in situations where the user does
not know that a particular input matrix is symmetric, or even that specialized
algorithms for symmetric matrices exist.

\section{Type system}

Julia uses run-time type tags to describe and differentiate objects by
their representation and behavior~\cite[Section 11.10, p. 142]{Pierce2002}.
We tend to use the term ``type'' instead of ``tag'', as most programmers seem comfortable
with this usage~\cite{Tratt2009,Kell2014}.

Tags are useful for both users and compilers for deciding what to do
to values, but incur overhead which increases the memory footprint of a data
item. This overhead motivates most dynamically typed languages to simplify and
minimize their tag systems.

Julia is unusual in allowing type tags with nested structure, forming nearly
arbitrary symbolic expressions. This has two immediate benefits: first,
it expands the criteria programmers have available for dispatching methods,
and second, provides richer information to the compiler.
Julia's type tags are designed to serve as elements of a lattice,
facilitating data flow analysis.

%
Julia types are first-class values, allowing programs to compute with them.
In technical computing it is unusually common to perform a non-trivial
computation to determine, for example, which data type to use for an operation.
Julia allows this to be done using ordinary code.

\subsection{Kinds of types}

In addition to tags, Julia has three other kinds of types, which together form
its full type lattice.
There are \emph{abstract} types, which form a nominal subtyping hierarchy.
Abstract types may serve as declared supertypes of tag types or of other abstract
types.
\emph{Union} types are used to form the least upper bound of any two types.
Finally, \emph{existential} types can be used to quantify over a type.

Tag types can have associated memory representations, corresponding to struct,
primitive, and array types familiar to the C language family.

\subsection{Subtyping}
\label{sec:subtyping}

Julia requires abstract and tag types to have exactly one
declared supertype, defaulting to \verb|Any| (i.e. $\top$) if not explicitly declared.
We conjecture that the subtype relation \verb|<:| is well-defined and
decidable, and that types are closed under meet ($\wedge$) and
join ($\vee$).




\subsection{Type parameters}
\label{sec:typeparameters}

Abstract and tag types can have one or more parameters. These
types are superficially similar to parametric types in existing
languages, but are actually intended simply for expressing information, rather
than implementing the formal theory of parametric polymorphism.

The following example, from Julia's standard library, describes a \verb|SubArray|
type that provides an indexed ``view'' into another array. Such array views
enable different indexing semantics to be overlaid on an existing array.
The \verb|SubArray| type
defines a \verb|Union| of what kinds of indexes may be used, then uses this
definition to define indexed views:

\begin{lstlisting}
  typealias RangeIndex Union(Int,Range{Int},
                             UnitRange{Int})

  type SubArray{T, N, A<:AbstractArray,
		I<:(RangeIndex...,)} <: AbstractArray{T,N}
      # definition body omitted
  end
\end{lstlisting}
\verb|SubArray| has parameters \verb|T| for an element type, \verb|N| for the
number of dimensions, \verb|A| for the underlying array type, and \verb|I|
for the tuple of indexes that describe which part of the underlying array
is viewed.
\verb|(RangeIndex...,)| denotes a tuple (essentially an immutable vector) of
any number of \verb|RangeIndex| objects.
The final \verb|<:| declares \verb|SubArray| to be a subtype of
\verb|AbstractArray|.

Julia's type system tries to hide type kinds from
the user. The identifier \verb|SubArray| by itself does not refer to a
type constructor that must be instantiated. Rather, it is a type
that serves as the supertype of all the \verb|SubArray| types. This allows
convenient shorthand for method signatures that are agnostic about type parameters.
It also makes it possible to add more type parameters in the future, without
being forced to update all code that uses that type.
This is achieved by making \verb|SubArray| an existential type with bounded
type variables.

When implementing \verb|SubArray| and client code using it, it is useful
to be able to dispatch on all of these parameters. For example, when
\verb|I| matches \verb|(UnitRange, RangeIndex...)| then the \verb|SubArray|
is contiguous in the leading dimension, and more efficient algorithms can
generally be used. Or, linear algebra code might want to restrict the
number of dimensions \verb|N| to 1 or 2.


\subsubsection{Variance}

The subtyping rules for parametric types require reasoning about variance,
i.e.\ the relation between subtype relations of the parametric types and
subtype relations of the parameters. The conventional wisdom is that type
safety allows covariance if components are read but not written, and
contravariance if they are written but not read~\cite{Castagna1995}. As type
parameters can represent the types of mutable fields, the only safe choice is
invariance. Thus Julia's parametric types are invariant.

\begin{figure}
	\centering
	\includegraphics[width=\columnwidth]{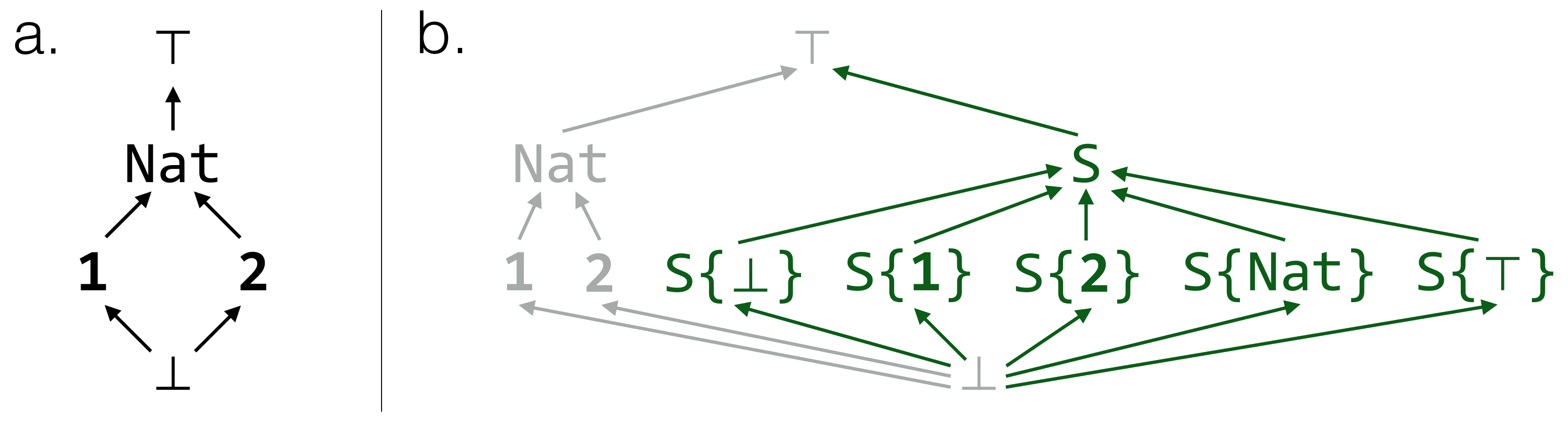}
	\caption{
		\textbf{a.} A simple lattice with bottom type $\bot$, top type
		$\top$, singleton types \textbf{1} and \textbf{2}, and their
		supertype \texttt{Nat}.
		\textbf{b.} The same lattice extended with a parametric type
		\texttt{S\{T\}} with no restriction on the type parameter
		\texttt{T}, showing that for each \texttt{T}, invariance
		requires that the corresponding type \texttt{S\{T\}} be a leaf
		type (i.e.\ instantiable), even if \texttt{T} itself is not a
		leaf type.}
	\label{fig:lattice}
\end{figure}

Parametric invariance has some subtle consequences for Julia's type system.
First, parametric types introduce many short, finite length chains into the
type lattice. Consider the simple type system of
Figure~\ref{fig:lattice}a, with two leaf (instantiable types) types \textbf{1}
and \textbf{2} representing singleton values 1 and 2 of the natural numbers
\verb|Nat|. A user can augment the lattice with a new parametric type
\verb|S{T}|. If there is no restriction whatsoever on the type parameter
\verb|T|, then there are 5 different parametric types of the form \verb|S{T}|.
Furthermore, each \verb|S{T}| has supertype \verb|S| by construction, and by
invariance of \verb|T| there are no values of the type parameters \verb|T| and
\verb|U| such that \verb|S{T}| is a subtype of \verb|S{U}|. Additionally, none
of the types \verb|S{T}| is a subtype or supertype of any of \textbf{1},
\textbf{2} or \verb|Nat|. Thus each \verb|S{T}| appears in exactly one finite
poset $\bot$ \verb|<: S{T} <: S <:| $\top$, and the new type lattice has the
structure shown in Figure~\ref{fig:lattice}b. Note that \verb|S{|$\top$\verb|}|
is a concrete type with type parameter $\top$ (\verb|Any|), while \verb|S{T}|
is a synonym for the abstract type \verb|S| where \verb|T| is a type variable
with lower bound $\bot$ and upper bound $\top$.

\subsection{Function types}

Julia is higher-order; functions can be passed to and returned from other functions.
However, the language currently does not have function (arrow) types.
The reason for this is that a Julia function need not have a useful canonical return type.
We could consider every function to have return type \verb|Any|, but this is not terribly revealing.
Return types based on Julia's type inference would not be predictable, as inference is heuristic and best-effort.
Future compiler improvements can make inferred types more specific, and this should not effect program behavior (only performance).

Julia does support \emph{nominal} function types --- a data type may implement
the \verb|call| generic function, allowing it to be used anywhere a function
can be used.

\subsection{Type conversion}

While Julia code can be written without
explicitly reasoning about types, the ability to do so is sometimes necessary for
understanding code performance issues. Julia provides the \verb|convert(T, x)|
function which converts a value \verb|x| to its corresponding representation in
the type \verb|T|. This is a generic function like any other, but since
\verb|T| is typically a compile-time constant, \verb|convert| this can serve
as a mechanism for limiting polymorphism in code where performance
considerations are important.

The idea of converting a value of type \verb|A| to type \verb|B|
does not naturally belong to one type or the other, which favors multiple
dispatch over classes. Conversion also benefits from open extension:
mathematical objects often have embeddings in multiple domains, not all of
which might be known or desired by the original author of some code. For
example, numbers can be embedded in matrices with diagonal matrices, but not
all users are likely to find this correspondence helpful.

\section{Generic functions and multimethods}

Mathematical thought is
naturally polymorphic. Multimethods are a natural mechanism for capturing such
polymorphism~\cite{Bezanson2014b,Chen2014}. Consider an operation as
fundamental as multiplication: an expression like \verb|a*b| can mean a matrix-matrix product, a matrix-vector product, or a scalar-vector product, to name just a few possibilities. A generic function system supporting
multimethods allows for the \verb|*| function to be polymorphic, expressing a
common metaphor for different kinds of multiplication which can be
disambiguated by the types of \verb|a| and \verb|b|. In contrast, languages
supporting only classes cannot capture the full extent of polymorphism in
\verb|*| in method dispatch: as classes inherently support only single dispatch
on the first argument, each method \verb|*| defined for each class \verb|a|
must contain different code blocks for each possible type of \verb|b|, thus in
practice requiring multiple dispatch to be emulated using virtual methods and
visitor patterns~\cite{designpatterns}. Furthermore, implementing binary methods can
require knowing the internal representation of both objects \verb|a| and
\verb|b|, especially for performance reasons~\cite{Bruce1995}. Such knowledge
fundamentally corrupts the abstraction of class-based encapsulation, as the
methods associated with \verb|a| must know implementation details of all
possible objects \verb|b| that \verb|a| may interact with.
In contrast, there is no abstraction leak associated with allowing a generic
function \verb|*| knowledge about the internal representations of the types it
works on.

Multiplication represented by \verb|*| can be extended, for
example, to multiplication between quaternions, or even to $N$-ary
matrix-matrix products, where associativity\footnote{Neglecting the lack of
exact associativity in some fields such as floating-point numbers.} allows
matrix-matrix products to be regrouped so as to minimize the total memory
footprint and operation count~\cite{Hu1984}.

The extensibility of Julia's generic functions and types allow users to
define new behaviors that intermix new types and new functions. The price we pay
for such flexibility, of course, is that dynamic function dispatch incurs
greater overhead: unlike in a single dispatch language, multiple lookups in
method tables may be necessary to determine which method is most
appropriate~\cite{Bruce1995}. Type inference is
useful for minimizing or even eliminating the
overhead associated with multiple dispatch.

Many of the use cases for multiple dispatch could potentially be addressed
by operator overloading as in C++. However C++ forces programmers to choose
which operations will use dynamic dispatch (via virtual methods), which will
use templates, and which will use function overloading. We conjecture that
this choice can be an unwelcome productivity drain. For example, the
syntactic difference between function calls and method calls could require
code to be rewritten when requirements change, or could lead to APIs that
are not consistent about when method calls are used.

\section{Type inference}
\label{sec:inference}

It is well known that type inference can be used to move tag manipulations
and method lookup to compile time, thus eliminating most overhead from
the execution of dynamically-typed code~\cite{Kaplan1977,Kaplan1980}.
Data flow type inference~\cite{Nielson2005,Khedker2009}
is especially useful in this context, as its
flow-sensitivity yields good type information even if the type of a program
variable is different at different points in the code.
Data flow analysis, particularly in the
forward direction, captures the human intuition of how programs work:
values start at the top and move through the program step by step.
Another advantage of this approach is that it is \emph{not} speculative:
it yields correct deductions about types that, in the best case, allow
overhead to be removed entirely. This property is important to technical
users, who need languages that can match the performance of C and Fortran.

Unfortunately, data flow type inference can be defeated
by dynamic language programs that are too ``type complex''. If the library
functions used might return too many different types, or there are too
many paths through user code, the resulting type information might not
be useful.

Julia was designed to help mitigate this problem. By encouraging
code to be written in small pieces labeled with type information (for
dispatch), it is easier for the compiler to rule out execution paths
that do not actually occur.

Type inference in Julia occurs after code is parsed, macro-expanded, and
lowered into a static single assignment (SSA) intermediate representation
(IR)~\cite{Alpern1988,Rosen1988} that facilitates data flow
analysis~\cite{Cousot1977,Cousot2000,Nielson2005} and is relatively
straightforward to map onto LLVM IR~\cite{Lattner2004}. Julia uses Mohnen's algorithm
for abstract interpretation~\cite{Cousot1992} which works directly on the SSA
IR~\cite{Mohnen2002}. The abstract semantics are described internally using
transfer functions (a.k.a.\ t-functions or flow functions), which approximate
program semantics by inferring possible output types based on the types of
the inputs.

In practice, the expressiveness of Julia's late binding semantics, combined
with the presence of potentially infinite types such as varargs tuples
\verb|(T...)|, complicate type inference.
Therefore practical type inference necessitates
widening heuristics, which reduce computational costs and guarantee termination
in the presence of recursive types~\cite{Cousot1992a}. Examples of such
heuristics include widening (pessimizing) type unions and tuple types which
exceed a certain length, and limiting the maximal depths of types
and tuples analyzed.

Julia provides language features to help users inspect the results of type
inference and specify additional type information where necessary to sharpen
the types of variables.

\begin{enumerate}

	\item The base library provides introspection functions like
	\verb|code_typed|, which allow users to inspect generated type
	annotations and avoid second-guessing the compiler's intentions.

	\item Variables can also be given explicit \textbf{type annotations};
	changing \verb|x = 0| to \verb|x::Float64 = 0| declares \verb|x| to be
	of type \verb|Float64| within the current scope, and all assignments
	\verb|x = _| implicitly call the type conversion function
	\verb|convert(Float64, _)|.

	\item Expressions can be given \textbf{type assertions}; changing
	\verb|x += y| to \verb|x += y :: Int| asserts that \verb|y| must be of
	type \verb|Int| or otherwise raise a run-time error.

\end{enumerate}
External packages like \package{TypeCheck.jl} and \package{Lint.jl} provide
further static analyses which are useful for detecting type-related issues.

\section{Applications in technical computing}

In this section, we describe how the type system, generic functions and type
inference interact in Julia code for scientific computations in base library
code as well as registered packages.

\subsection{Type promotion}

The type system and generic function system allow for type promotion rules to
be specified in the base library rather than be hard-coded into a given
implementation~\cite{Bezanson2012a}. For example, simple arithmetic functions
on general numeric types in Julia's base library contain methods of the form:

\begin{lstlisting}
+(x::Number, y::Number) = +(promote(x,y)...)
*(x::Number, y::Number) = *(promote(x,y)...)
-(x::Number, y::Number) = -(promote(x,y)...)
/(x::Number, y::Number) = /(promote(x,y)...)
^(x::Number, y::Number) = ^(promote(x,y)...)
\end{lstlisting}

For example, an operation like \verb|2 * 3.4| is evaluated under the hood as
follows:

\begin{lstlisting}
*(2, 3.4) = *(promote(2, 3.4)...)
          = *(convert(promote_type(Int,Float64),2),
	      convert(promote_type(Int,Float64),3.4))
          = *(convert(Float64,2), convert(Float64,3.4))
	  = *(2.0, 3.4)
	  = 6.8
\end{lstlisting}
where \verb|promote(x, y)| promotes the values \verb|x| and \verb|y| to
numerically equivalent values of a common supertype, and
\verb|promote_type(S,T)| computes a common supertype of \verb|S| and \verb|T|.
The latter calls a custom promotion rule, if defined, and defaults otherwise to
the join \verb|S|$\vee$\verb|T|.

Type promotion allows for different functions and methods to share a common and
consistent logic for polymorphism. The implementation of type promotion
leverages the type system to allow for greater code reuse across different
methods and functions. Furthermore, it is a part of the Julia language which
can be built entirely from other language constructs.

\subsection{Numerical linear algebra library}

Designing a general-purpose linear algebra library involves several different
layers of complexity, and has been described as implementing the following
meta-program~\cite{Demmel2007}:

{\small
\begin{verbatim}
(1) for all linear algebra problems
    (linear systems, eigenproblems, ...)
(2)  for all matrix types
     (general, symmetric, banded, ...)
(3)   for all data types (real, complex,
      single, double, higher precision)
(4)    for all machine architectures and
       communication topologies
(5)     for all programming interfaces
(6)      provide the best algorithm(s) available
         in terms of performance and accuracy
         (``algorithms'' is plural because
         sometimes no single one is always best)
\end{verbatim}
}
The six-tiered hierarchy neatly delineates how the basic collection of linear
algebra problems (1) have to be specialized by data representation (2--3) and
machine details (4), which are then used to decide which specific algorithms
(5--6) to use.

Many systems provide optimized implementations of standard libraries for
numerical linear algebra like BLAS (Basic Linear Algebra Subprograms) and
LAPACK (Linear Algebra PACKage)~\cite{lapack}. Computational routines can be
customized for individual microarchitectures and can reach a large fraction of
theoretical peak FLOPS. However, these libraries inherit archaic Fortran 77
interfaces and hence tend to restrict routine names to six letters or shorter.
When combined with the lack of polymorphism in Fortran, the names are terse and
cryptic to nonexperts: a typical routine like \verb|DSTEV| solves the
eigenvalue problem (\verb|EV|) for symmetric tridiagonal matrices (\verb|ST|)
with double precision floating point entries (\verb|D|), and furthermore takes
eight positional arguments specifying the inputs, outputs, computation mode,
and scratch variables. The lack of polymorphism results in redundancy due to
lack of code reuse, which hinders the implementation of new algorithms (which
have to be reimplemented for each level of floating-point precision) and new
routines for such as mixed-precision and quad precision routines (which must
implement all the existing algorithms).

The code redundancy problem is largely ameliorated with the combination of type
polymorphism and dynamic multiple dispatch. The six-tiered hierarchy above maps
naturally onto different language constructs in Julia as follows:

\vspace{12pt}
\begin{tabular}{c l l}
	\hline
	Tier & Description & Language construct \\ \hline
	1 & linear algebra problems & generic function \\
	2 & matrix types & parametric types \\
	3 & data types & type parameter \\
	4 & machine architectures & method body \\
	5 & programming interfaces & generic function \\
	6 & (poly)algorithms & method body \\ \hline
\end{tabular}
\vspace{12pt}

The generic function system allows for fast specializations and generic
fall-backs to coexist, thus allowing for speed when possible and flexibility
otherwise. For example, Julia provides generic fall-back routines to do matrix
computations over arbitrary fields of element types, providing the ability to
compute on numeric types which are not mappable to hardware floating point
types. This can be useful to perform matrix computations in exact rational
arithmetic or software-emulated higher precision floating point arithmetic to
verify the implementations of algorithms or to detect the possibility of
numerical instability associated with roundoff errors. These general purpose
routines coexist with BLAS and LAPACK wrappers, thus allowing dispatch onto
performant code when available, and general code otherwise. User code can be
written that will work regardless of element type (Tier 3), and can be tuned
for performance later.

\subsection{Generic linear algebra}

Code for technical computing often sacrifices abstraction for performance,
and is less expressive as a result. In contrast, mathematical ideas are
inherently polymorphic and amenable to abstraction. Consider a simple example
like multiplication, represented in Julia by the \lstinline|*| operator.
\lstinline|*| is a generic function in Julia, and has specialized methods for
many different multiplications, such as scalar--scalar products, scalar--vector
products, and matrix--vector products. Expressing all these operations using
the same generic function captures the common metaphor of multiplication.

\subsection{Bilinear forms}

Julia allows user code to extend the \lstinline|*| operator, which can be
useful for more specialized products. One such example is bilinear forms, which
are vector--matrix--vector products of the form
\begin{equation}
\gamma = v^\prime M w = \sum_{ij} v_i M_{ij} w_j
\end{equation}
This bilinear form can be expressed in Julia code as
\begin{lstlisting}
function (*){T}(v::AbstractVector{T},
                M::AbstractMatrix{T},
                w::AbstractVector{T})
    if !(size(M,1) == length(v) &&
         size(M,2) == length(w))
        throw(BoundsError())
    end
    $\gamma$ = zero(T)
    for i = 1:size(M,1), j = 1:size(M,2)
        $\gamma$ += v[i] * M[i,j] * w[j]
    end
    return $\gamma$
end
\end{lstlisting}
The newly defined method can be called in an expression like
\lstinline|v * M * w|, which is parsed and desugared into an ordinary function
call \lstinline|*(v, M, w)|. This method takes advantage of the result being a
scalar to avoid allocating intermediate vector quantities, which would be
necessary if the products were evaluated pairwise like in $v^\prime(Mw)$ and
$(v^\prime M) w$. Avoiding memory allocation reduces the number of
heap-allocated objects and produces less garbage, both of which are important
for performance considerations.

The method signature above demonstrates two kinds of polymorphism in Julia.
First, both \lstinline|AbstractVector| and \lstinline|AbstractMatrix| are
abstract types, which are declared supertypes of concrete types. Examples of
subtypes of \lstinline|AbstractMatrix| include \lstinline|Matrix| (dense
two-dimensional arrays) and \lstinline|SparseMatrixCSC| (sparse matrices stored
in so-called compressed sparse column format). Thus the method above is defined
equally for arrays of the appropriate ranks, be they dense, sparse, or even
distributed.  Second, \lstinline|T| defines a type parameter that is common to
\lstinline|v|, \lstinline|M| and \lstinline|w|. In this instance, \lstinline|T|
describes the type of element stored in the \lstinline|AbstractVector| or
\lstinline|AbstractMatrix|, and the \lstinline|{T}(...{T}...{T})| syntax
defines a family of methods where \lstinline|T| is the same for all three
arguments. The type parameter \lstinline|T| can also used in the function body;
here, it is used to initialize a zero value of the appropriate type for
$\gamma$.

The initialization statement \lstinline|$\gamma$ = zero(T)| requests a zero
element of the appropriate type.
In Julia this is important because using a zero value of a different type can
cause a different method to be called.
If \lstinline|T| is any concrete 
type, e.g.\ \lstinline|Float64| (64-bit floating point real numbers), then
Julia's just-in-time compiler can analyze the code statically to remove type
checks.  For example, in the method with signature
\lstinline|*{Float64}(v::AbstractVector{Float64}, M::AbstractMatrix{Float64}, w::AbstractVector{Float64})|,
the indexing operations on $v$, $M$ and $w$ always return \lstinline|Float64|
scalars. Furthermore, forward data flow analysis allows the type of $\gamma$
to be inferred as \lstinline|Float64| also, since floating point numbers are
closed under addition and multiplication. Hence, type checks and method
dispatch for functions like \lstinline|+| and \lstinline|*| within the function
body can be resolved statically and eliminated from run time code, allowing
fast code to be generated.

Were we to replace the initialization with the similar-looking
\lstinline|$\gamma$ = 0|, we would have instead a \textbf{type instability}
when \lstinline|T = Float64|. Because $\gamma$ is initialized to an
\lstinline|Int| (native machine integer) and it is incremented zero or more
times by a \lstinline|Float64| in the \lstinline|for| loop, the type of
the variable can change at run time.
In fact, the result type will depend on the size of the input array $M$,
which is only known at run time.
Julia allows this behavior, and our
compiler infers the type of $\gamma$ to be \lstinline|Union(Int,Float64)|,
which is the least upper bound on the actual type of $\gamma$ at run time.
As a result, not all type
checks and method dispatches can be hoisted out of the loop body, resulting
in slower execution.

\subsection{Matrix equivalences}

Matrix equivalences are another example of specialized product that users may
want. Two $n\times n$ matrices $A$ and $B$ are considered equivalent if there
exist invertible matrices $V$ and $W$ such that $B = V * A * W^\prime$.
Oftentimes, equivalence relations are considered between a given matrix $B$ and
another matrix $A$ with special structure, and the transformation
$(W^\prime)^{-1} \cdot V^{-1}$ can be thought of as changing the bases of the
rows and columns of $B$ to uncover the special structure buried within as $A$.
Matrices with special structure are ubiquitous in numerical linear algebra. One
example is rank-$k$ matrices, which can be written in the outer product form $A
= X Y^\prime$ where $X$ and $Y$ each have $k$ columns. Rank-$k$ matrices may be
reified as dense two-dimensional arrays, but the matrix--matrix--matrix product
$V * A * W^\prime$ would take $O(n^3)$ time to compute. Instead, when $k \ll
n$, the product be computed more efficiently in $O(kn^2)$ time, since
\begin{equation}
V A W^\prime = V (X Y^\prime) W^\prime = (V X) (W Y)^\prime
\end{equation}
and the result is again a rank-$k$ matrix. Furthermore, we can avoid
constructing $W^\prime$, the transpose of $W$, explicitly. Therefore in some
cases, it is sensible to store $A$ as the two matrices $X$ and $Y$ separately,
rather than as a reified 2D array.

Julia allows users to encapsulate $X$ and $Y$ within a specialized type:
\begin{lstlisting}
type OuterProduct{T}
	X :: Matrix{T}
	Y :: Matrix{T}
end
\end{lstlisting}
Defining the new \lstinline|OuterProduct| type has the advantage of grouping
together objects that belong together, but also enables dispatch based on the
new type itself. We can now write a new method for \lstinline|*|:
\begin{lstlisting}
*(V, M::OuterProduct, W) = OuterProduct(V*M.X, W*M.Y)
\end{lstlisting}
This method definition uses a convenient one-line syntax for short definitions
instead of the \lstinline|function ... end| block. This method also does not
annotate \lstinline|V| or \lstinline|W| with types, and so they are considered to
be of the top type $\top$ (\lstinline|Any| in Julia). This method may be called
with any $V$ and $W$ which support premultiplication: so long as
\lstinline|V*M.X| and \lstinline|W*M.Y| are defined and produce matrices of the
same type, then the code will run without errors (``duck typing'').
This flexibility is
convenient since $V$ and $W$ can now be scalars or matrixlike objects which
themselves have special structures, or even more generally could represent
linear maps that are not stored explicitly, but rather defined implicitly
through their actions when multiplying a \lstinline|Matrix| on the left.

The preceding method shows that Julia does not require all method arguments to
have explicit type annotations. Instead, dynamic multiple dispatch allows the
argument types to be determined from the arguments at run time.
Julia's just-in-time compiler will be invoked when the method is first
called, so that static analyses can be performed.

We can now proceed to define a new type and method

\begin{lstlisting}
type RowPermutation
	p::Vector{Int}
end

*($\Pi$::RowPermutation, M::Matrix) = M[$\Pi$.p, :]
\end{lstlisting}
whose action can be thought of as multiplying by a permutation matrix on the
left, resulting in a version of $M$ with the rows permuted. Now, the following
user code
\begin{lstlisting}
n = 10
k = 2
X = randn(n, k) #Random matrix of Float64s
M = OuterProduct(X, X)
p = randperm(n) #Random permutation of length n
$\Pi$ = RowPermutation(p)
M2 = $\Pi$ * M * $\Pi$
\end{lstlisting}
will dispatch on the appropriate methods of \lstinline|*| to produce the same
result \lstinline|M2| as
\begin{lstlisting}
M2 = OuterProduct(M.X[p, :], M.Y[p, :])
\end{lstlisting}
In other words, the specialized method
\lstinline|*(::RowPermutation, ::OuterProduct{Float64}, ::RowPermutation)|
is compiled only when it is first invoked in the creation of
\lstinline|M2|, since it follows from composing the method defined with
signature \lstinline|*(::Any, ::OuterProduct, ::Any)| with the argument tuple
of type \lstinline|(RowPermutation, OuterProduct{Float64}, RowPermutation)|.

\subsection{Matrix factorization types}

Julia's base linear algebra library also provides extensive support for matrix
factorizations, which are indispensable for reasoning about the
interdependencies between matrices with special properties and the numerical
algorithms that they enable~\cite{Golub2013}. Many algorithms for numerical
linear algebra involve interconversions between general matrices and similar
matrices with special matrix symmetries. For many purposes, it is convenient to
reason about the resulting special matrix, together with the matrix performing
the transformation, as a single mathematical object rather than two separate
matrices. Such an object represents a matrix factorization, and is essentially
a different data structure that can represent the same content as a matrix
represented as an ordinary two-dimensional array.

The exact matrix factorization object relevant for a given linear algebra
problem depends on the symmetries of the starting matrix and also the
underlying field of matrix elements (i.e.\ whether the matrix contains real
numbers, complex numbers, or something else). In some use cases, these
properties may be known by the user as part of the problem specification, and
in other cases they may be unknown. Some properties, like whether a
matrix is triangular, can be deduced by inspecting the matrix elements for
$O(N^2)$ cost. Others, like whether a matrix is positive definite, require an
$O(N^3)$ computation in the general case, and is most efficiently determined by
attempting to compute the Cholesky factorization.

The resulting algorithm for determining a useful matrix factorization has to
capture the interplay between allowing the user to specify additional matrix
properties and attempting to automatically detect useful properties when the
additional cost of doing so is not prohibitive. The general case is implemented
in Julia's base library by the \verb|factorize| function, and typifies the
complexity associated with numerical codes:

\begin{lstlisting}
function factorize{T}(A::Matrix{T})
	m, n = size(A)
	if m != n
		# A is rectangular
		# Can the result of a QR factorization be represented
		# using a floating point type supported by BLAS?
		BlasFloat = Union(Float32,Float64,Complex64,Complex128)
		rt = zero(T) / sqrt(zero(T) + zero(T))
		can_use_BlasFloat = isa(rt, BlasFloat)
		# Factorize into pivoted QR form where possible,
		# otherwise compute (unpivoted) QR form
		return qrfact(A, pivot=can_use_BlasFloat)
	end
	# A is square
	# The factorization of a 1x1 matrix is just itself
	m == 1 && return A[1]
	utri = istriu(A)          # upper triangular?
	utri1 = ishessenbergu(A)  # upper Hessenberg?
	sym = issym(A)
	herm = ishermitian(A)
	ltri = istril(A)          # lower triangular?
	ltri1 = ishessenbergl(A)  # lower Hessenberg?
	if ltri && utri
		return Diagonal(A)
	elseif ltri && utri1
		return Bidiagonal(diag(A), diag(A, -1), false)
	elseif ltri
		return Triangular(A, :L)
	elseif ltri1 && utri
		return Bidiagonal(diag(A), diag(A, 1), true)
	elseif ltri1 && utri1
		if (herm && (T <: Complex)) || sym
			M = SymTridiagonal(diag(A), diag(A, -1))
			# _may_ be factorizable into
			# LDL' Cholesky form
			try
				return ldltfact!(M)
			end
		end
		# Factorize into tridiagonal LU form
		M = Tridiagonal(diag(A,-1), diag(A), diag(A,1))
		return lufact!(M)
	elseif utri
		return Triangular(A, :U)
	elseif herm
		# try to factorizable into Cholesky form
		try
			return cholfact(A)
		end
		# else use general Hermitian factorization
		return factorize(Hermitian(A))
	elseif sym
		# Use general Symmetric factorization
		return factorize(Symmetric(A))
	else
		# A is square but has no other symmetries
		# Factorize into LU form
		return lufact(A)
	end
end

# Factorize a Hermitian or Symmetric matrix into
# Bunch-Kaufman form as computed by bkfact
typealias HermOrSym Union(Hermitian,Symmetric)

factorize(A::HermOrSym) =
    bkfact(A.data, symbol(A.uplo), issym(A))
\end{lstlisting}

The \verb|factorize| function contains a few interesting features
designed to capture the highly dynamic nature of this computation:

\begin{enumerate}
\item The beginning checks for the ``easy'' properties, i.e.  those that can be
	computed cheaply at run-time by inspecting the matrix elements. The
	presence of one or more of these properties allow the input matrix
	\verb|A| to be classified into several special cases.

\item For many of these special cases (like \verb|Diagonal|), \verb|A| is
	explicitly converted to a special matrix type which allows dispatch on
	efficient specializations of linear algebraic operations that users may
	choose to perform later, such as matrix multiplication.

\item For other special cases (like \verb|Tridiagonal|), it is useful to
	attempt a run-time check for the ``hard'' properties, like positive
	definiteness, which require nontrivial computation. Computing the
	Cholesky factorization serves both as an efficient run-time check to
	detect positive definiteness, whose failure indicates lack thereof, and
	whose success yields a useful \verb|Cholesky| matrix factorization
	object which is useful for further computations such as solving linear
	equations or computing singular value decompositions (SVDs).

\item The \verb|Symmetric| and \verb|Hermitian| cases use control flow
	enabled by dynamic multimethods, as it is common for users to know
	whether the matrices they are working with have these properties.
	Using multimethods allows users to elide all the run-time checks
	in the generic method, skipping directly to the appropriate specialized
	method.

\item The base case contains a nontrivial runtime check to determine if the
	element type of \verb|A{T}| can be represented as a
	hardware-representable real or complex floating point number. The
	\verb|canuseBlasFloat| variable computes the type resulting from
	computations of the form $x / \sqrt{x + x}$, and checks that it is a
	subtype of \verb|BlasFloat|. The reason for this check is that there
	are two variants of the QR factorization: one pivoted, the other not.

\end{enumerate}

In theory, it is possible to combine these structured matrix types under a
tagged union, e.g.\ in an ML-family language. However this would be
less convenient. The problem is that in most contexts,
users are happy to have these objects separated by the type system, but
certain functions like \code{factorize} wish to combine them. It should be
possible to pass the result of such a function directly to an existing
routine that expects a particular matrix structure, without needing to
interpose case analysis to handle a tagged union.

It is also instructive to look more carefully at the methods for \verb|bkfact|,
which is but one of the several factorizations computed in \verb|factorize|:

\begin{lstlisting}
# Compute Bunch-Kaufman factorization,
bkfact{T<:BlasFloat}(A::StridedMatrix{T},
                     uplo::Symbol=:U,
                     sym::Bool=issym(A)) =
    bkfact!(copy(A),uplo,sym)

function bkfact{T}(A::StridedMatrix{T},
                   uplo::Symbol=:U,
                   sym::Bool=issym(A))
    Typ = promote_type(Float32,
                       typeof(sqrt(one(T))))
    bkfact!(Matrix{Typ}(A),uplo,symmetric)
end

# Compute Bunch-Kaufman factorization in-place
# call LAPACK SSYTRF/DSYTRF where possible
function bkfact!{T<:BlasReal}(A::StridedMatrix{T},
		              uplo::Symbol=:U,
			      sym::Bool=issym(A))
    if sym
        error("The Bunch-Kaufman decomposition ",
	      "is only valid for symmetric matrices")
    end
    LD, ipiv = LAPACK.sytrf!(char_uplo(uplo),A)
    BunchKaufman(LD,ipiv,char_uplo(uplo),symmetric)
end

function bkfact!{T<:BlasComplex}(A::StridedMatrix{T},
	    			 uplo::Symbol=:U,
				 sym::Bool=issym(A))
    if sym
        LD,ipiv = LAPACK.sytrf!(char_uplo(uplo),A)
    else
        LD,ipiv = LAPACK.hetrf!(char_uplo(uplo),A)
    end
    BunchKaufman(LD,ipiv,char_uplo(uplo),sym)
end
\end{lstlisting}

The Bunch-Kaufman routines are implemented in two functions: \verb|bkfact|,
which allocates new memory for the answer, and \verb|bkfact!|, which mutates
the matrix in place to save memory. Reasoning about memory use is critical in
numerical linear algebra applications as the matrices may be large.
Reusing allocated memory for matrices also avoids unnecessary copies of data to
be made, potentially minimizing memory accesses and reducing the need to
trigger garbage collection. The base library thus provides the latter for users
who need to reason about memory usage, and the former for users who do not.

The syntax \verb|{T<:BlasReal}| in a method definition implicitly wraps the
signature in a \verb|UnionAll| type. As a result, the method matches all
matrices whose elements are of a type supported by BLAS.

\section{Case study: completely pivoted LU}

Linear algebra is ubiquitous in technical computing applications. At the same
time, the implementation of linear algebra libraries is generally considered a
difficult problem best left to the experts. A popular reference book for
numerical methods famously wrote, for example, that ``the solution of
eigensystems... is one of the few subjects covered in this book for which we do
\textit{not} recommend that you avoid canned routines''~\cite[Section 11.0, p.
461]{Press1992}. While much effort has been invested in making
numerical linear algebra libraries
fast~\cite{lapack,Gunnels2001,OpenBLAS,VanZee2013}, one nevertheless will
occasionally need an algorithm that is not implemented in a
standard linear algebra library.

One such nonstandard algorithm is the completely pivoted LU factorization. This
algorithm is not implemented in standard linear algebra libraries in
LAPACK~\cite{lapack}, as the conventional wisdom is the gains in numerical
stability in complete pivoting is not generally worth the extra effort over
other variants such as partial pivoting~\cite{Golub2013}.  Nevertheless, users
may want complete pivoting for comparison with other algorithms for a
particular use case.

In this section, we compare implementations and performance of this
algorithm in Julia with other high level languages that are commonly used for
technical computing: MATLAB, Octave, Python/NumPy, and R.

\subsection{Na\"ive textbook implementation}

Algorithm~\ref{alg:lucompletepiv} presents the textbook description of the LU
factorization with complete pivoting~\cite[Algorithm 3.4.3 (Outer Product LU
with Complete Pivoting), p. 132]{Golub2013}, and below it a direct translation
into a na\"ive Julia implementation. This algorithm is presented in MATLAB-like
pseudocode, and contains a mixture of scalar \lstinline|for| loops and
MATLAB-style vectorized indexing operations that describe various subarrays of
the input matrix $A$. Additionally, there is a description for the subproblem
of finding the next pivot at the start of the loop. Furthermore, the pseudocode
uses the $\leftrightarrow$ operation, denoting swaps of various rows and
columns of $A$. To translate the pivot finding subproblem into Julia, we used
the built-in \lstinline|indmax| function to find the linear index of the value
of the subarray \lstinline|A[k:n, k:n]| which has the largest magnitude, then
used the \lstinline|ind2sub| function to convert the linear index to a tuple
index. The $\leftrightarrow$ operator was implemented using vectorized indexing
operations, as is standard practice in high level languages like MATLAB.

\begin{algorithm}

\caption{Top: Textbook pseudocode describing the $LU$ factorization with
complete pivoting~\cite[Algorithm 3.4.3 (Outer Product LU with Complete
Pivoting), p. 132]{Golub2013}. The matrix $A$ is overwritten in-place with the
$LU$ factors, with $rowpiv$ and $colpiv$ containing the row and column pivots
respectively.
Bottom: An implementation of $LU$ factorization with complete pivoting in Julia,
which returns the result as a tuple. The ! at the end of the function name is
convention for a function with side effects (in this case, mutating $A$).
Unicode characters such as Greek letters and the $\ne$ operator are allowed in
Julia code, allowing for close notational correspondence with the textbook
description of the algorithm.}
\label{alg:lucompletepiv}

\begin{algorithmic}
\For {$k = 1:n - 1$}

    Determine $\mu, \lambda$ where $k \le \mu \le n$,  $k \le \lambda \le n$, so

    $\quad\left|A(\mu, \lambda)\right| = \max\{ \left|A(i, j)\right| : i=k:n, j=k:n \}$

    $rowpiv(k) = \mu$

    $A(k, 1:n) \leftrightarrow A(\mu, 1:n)$

    $colpiv(k) = \lambda$

    $A(1:n, k) \leftrightarrow A(1:n, \lambda)$

    \If{$A(k,k) \ne 0$}

        $\rho = k+1:n$

        $A(\rho, k) = A(\rho, k)/A(k, k)$

        $A(\rho, \rho) = A(\rho, \rho) - A(\rho, k) A(k, \rho)$
    \EndIf
\EndFor
\end{algorithmic}

\hrulefill

\begin{lstlisting}
function lucompletepiv!(A)
	n=size(A, 1)
	rowpiv=zeros(Int, n-1)
	colpiv=zeros(Int, n-1)
	for k=1:n-1
		Asub = abs(A[k:n, k:n]) #Search for next pivot
		$\mu$, $\lambda$ = ind2sub(size(Asub), indmax(Asub))
		$\mu$ += k-1; $\lambda$ += k-1
		rowpiv[k] = $\mu$
		A[[k, $\mu$], 1:n] = A[[$\mu$, k], 1:n]
		colpiv[k] = $\lambda$
		A[1:n, [k, $\lambda$]] = A[1:n, [$\lambda$, k]]
		if A[k,k] $\ne$ 0
			$\rho$ = k+1:n
			A[$\rho$, k] = A[$\rho$, k]/A[k, k]
			A[$\rho$, $\rho$] = A[$\rho$, $\rho$] - A[$\rho$, k] * A[k, $\rho$]
		end
	end
	return (A, rowpiv, colpiv)
end
\end{lstlisting}

\end{algorithm}

\begin{figure}
	\includegraphics[width=\columnwidth]{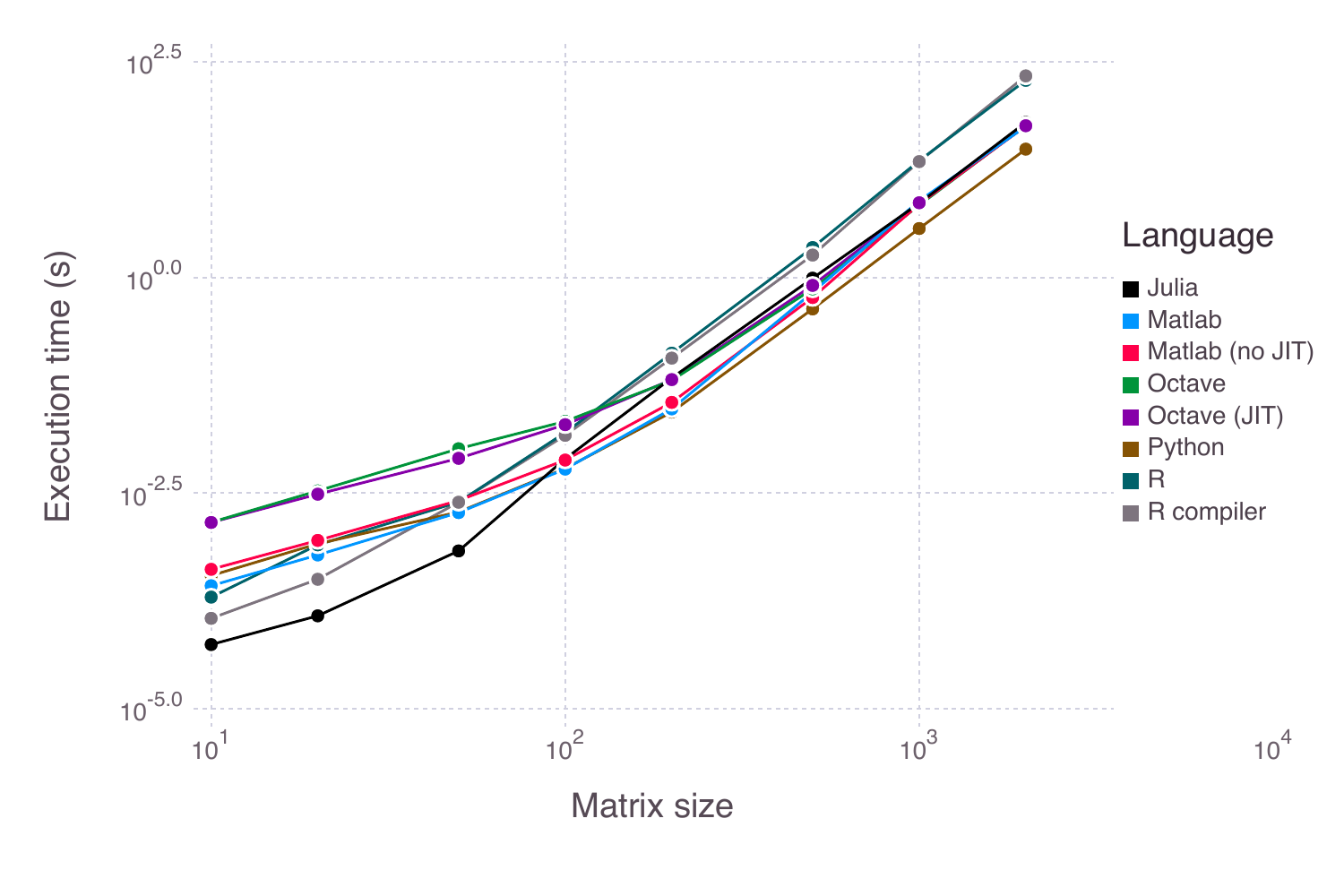}
	\caption{Scaling behavior of na\"ive implementations of the completely
	pivoted $LU$ algorithm on $N\times N$ random matrices in Julia, MATLAB,
	Octave, Python/NumPy, and R. By default, MATLAB's JIT compiler is on,
	whereas Octave's and R's are off. Julia code is listed in
	Algorithm~\ref{alg:lucompletepiv}.
}
	\label{fig:scaling}
\end{figure}

\begin{figure}
	\includegraphics[width=\columnwidth]{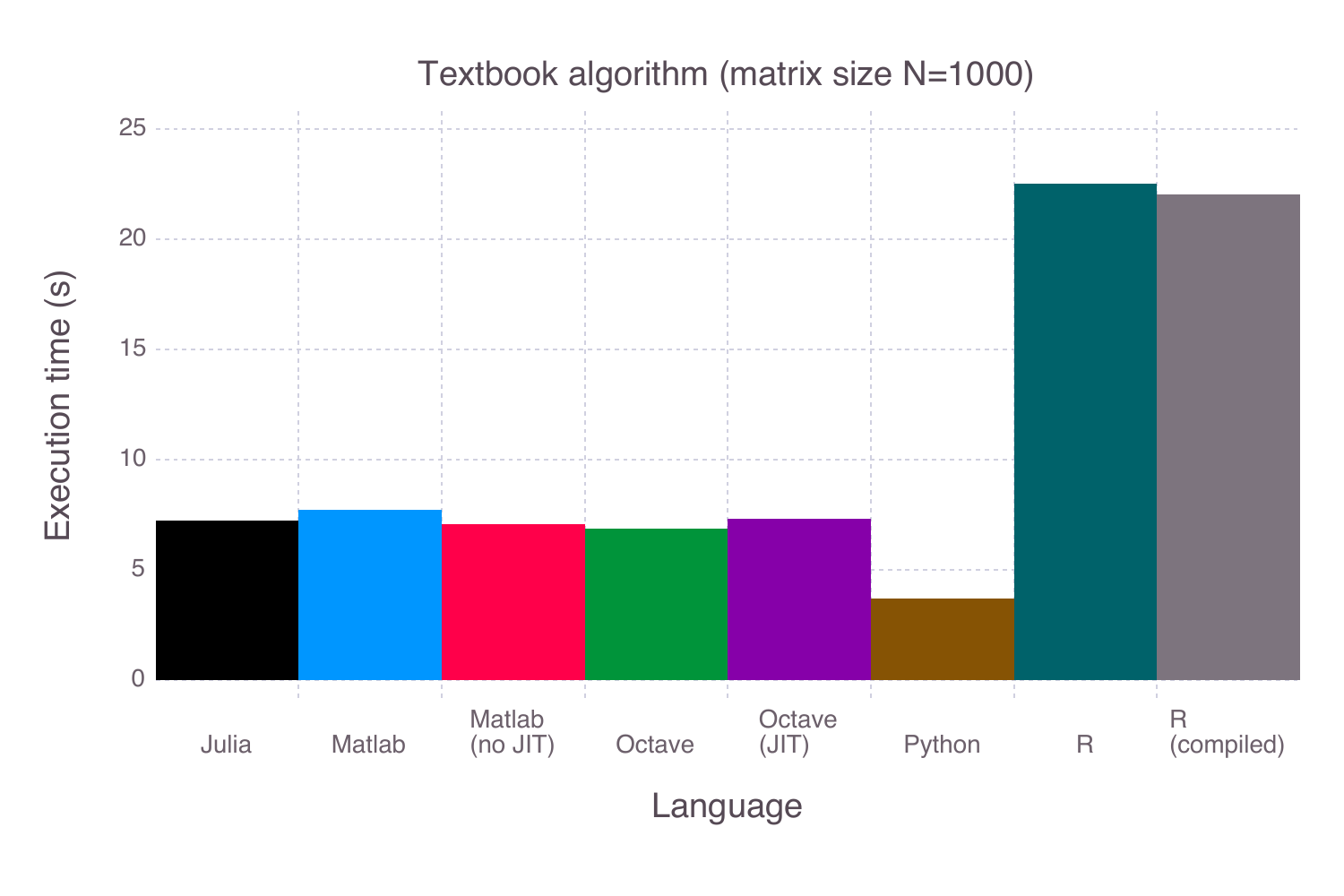}
	\caption{Execution times of na\"ive implementations of the
	completely pivoted $LU$ algorithm on a $1000\times1000$ random matrix
	in Julia, MATLAB, Octave, Python/NumPy, and R. See Figure~\ref{fig:scaling}
	for further details.}
	\label{fig:naivelangs}
\end{figure}

For comparison purposes, we also wrote na\"ive implementations in other high
level dynamics languages which are popular for technical computing. Here, we
considered MATLAB~\cite{matlab}, Octave~\cite{octave},
Python~\cite{python}/NumPy~\cite{numpy}, and R~\cite{rlang} (whose codes are
available in the Appendix). The codes were executed on a late 2013 MacBook Pro
running OS X
Yosemite 10.10.2, with Julia 0.4-dev+3970, MATLAB R2014b, Octave 3.8.1, Python
3.4.3 with NumPy 1.9.2 from the Anaconda 2.1.0 distribution, and R 3.1.3. Where
possible, we also tried to run variants with and without JIT compilation. In
MATLAB, the JIT compiler is on by default, but can be turned off with the
command \lstinline|feature accel off|. Octave's JIT compiler is experimental
and off by default, but can be enable with a command line switch. R provides a
JIT compiler in the \lstinline|compiler| library package. We do not have
JIT-compiled results for Python, as at this time of writing, neither PyPy
2.5.0~\cite{Bolz2009} nor Numba 0.17.0 was able to compile the code.\footnote{
The specialized fork of NumPy required to run on PyPy 2.5.0 did not build
successfully on neither Python 2.7.9 nor 3.4.3 on OSX. Numba 0.17.0, with the
\lstinline|@jit(nopython=True)| decorator, threw a
\lstinline|NotImplementedError| exception.}
The results are summarized in Figure~\ref{fig:scaling}, which shows the
near-perfect $O(n^3)$ scaling of the algorithm in each implementation, as well
as in Figure~\ref{fig:naivelangs}, which shows the execution times across the
different languages for a $1000 \times 1000$ matrix of \lstinline|Float64|s
with randomly sampled standard Gaussians.

The results show that the na\"ive Julia implementation performs favorably with
the implementations in Matlab and Octave, all of which are significantly faster
than R's implementation. Python's implementation is somewhat faster, owing to
the rank-1 update step (of \lstinline|A[$\rho$, $\rho$]|)
being done in place. Turning on or off the JIT did not significantly change the
execution times for Matlab, Octave and R.

\subsection{LU decomposition on arbitrary numeric types}

One major advantage to writing the $LU$ factorization code in pure Julia is
that \lstinline|lucompletepiv!| can run on any \lstinline|Matrix{T}|. So long
as the underlying element type \lstinline|T| is closed under the basic
arithmetic operations \lstinline|+|, \lstinline|-|, \lstinline|*|,
\lstinline|/|, \lstinline|abs|, and \lstinline|max|, the algorithm will run
just as it did on \lstinline|Float64| numbers. For example, one could compute
the completely pivoted $LU$ factorization on matrices of fixed point numbers
(provided by the \package{FixedPointNumbers.jl} package), or matrices of
rational numbers (built in \lstinline|Rational| types).

\begin{lstlisting}
#Initialize 1000x1000 matrix of 32-bit fixed point 1s
#with 14 fractional bits
using FixedPointNumbers
B = ones(Fixed32{14}, 1000, 1000)
lucompletepiv!(B)

#Initialize 1000x1000 matrix of unit rational numbers
#with 64-bit integer numerators and denominators
C = ones(Rational{Int64}, 1000, 1000)
lucompletepiv!(C)
\end{lstlisting}

Other interesting examples include the dual and hyperdual numbers provided by
the \package{DualNumbers.jl} and \package{HyperDualNumbers.jl} packages
respectively. Both types of numbers are used for forward mode automatic
differentiation, and can be used with \lstinline|lucompletepiv!| to take
derivatives of the $LU$ factorization itself. Computations over such arbitrary
numeric types would be difficult, if not impossible, in the other languages,
without reimplementing the basic algorithm \lstinline|lucompletepiv!| over and
over again.

While of course it is possible to build in more numeric types to the base
library of any programming language, the approach shown here is more general
by virtue of being extensible by users. Other such numberlike quantities
include colors (in \package{Color.jl}), unitful quantities (in
\package{SIUnits.jl}), interval arithmetic (in \package{ValidatedNumerics.jl}),
finite fields, dates and times (in \lstinline|Base.Dates|), quaternions and
octonions (in \package{Quaternions.jl}, extended precision floating point (in
\package{DoubleDouble.jl}), DNA nucleotides (in \package{BioSeq.jl}), and many
more.

\subsection{Improving the performance of a na\"ive implementation}
One of the reasons why high level languages are slow is that it allows the programmer to express algorithms in terms of array operations. These languages have a limited ability to optimize array expressions and consequently many temporary arrays are allocated during execution of a program.

In languages such as Fortran an C, similar algorithms are usually written without allocating temporary arrays. Some workspace might be required by the algorithm, but memory is then typically allocated once. This makes it possible to compile the code into efficient machine code that is typically much faster than what is possible for higher level array oriented languages.

In Julia, it is possible to express algorithms in terms of array operations, but it is also possible to avoid array allocations and thereby have the compiler optimizing the code to efficient machine code. Hence, a first step in optimizing Julia code is to find the lines during a loop that allocates temporary arrays.

The first line in the loop body makes two unnecessary array copies. The lines that flip columns and rows also allocate temporary arrays which can be avoided and allocations are also made for the scaling and rank one update in the last two lines of the \texttt{if} statement. By writing small auxiliary functions and expanding the scaling and rank one update as loops, it is possible to reduce the number of temporary allocations significantly.

For a square matrix of size 1000, a profiling of the na\"ive implementation reveals that it  allocates more than 12 GB of memory and runs in 4.15 seconds. With the changes mentioned in last paragraph the memory allocation is only 7 MB and the running time reduces to 0.75 seconds. Typically, the avoidance of array allocation amounts to the largest share of speedup when optimizing Julia code, but it is possible to achieve a further speed improvement by annotating the code with two macros that turn off bounds checking on array indexing and allows the compiler to use SIMD registers when possible. This final optimization reduces the running time to 0.4 seconds.

In many cases it is not desired to write out array operations as loops, but it is convenient that this optimization is possible without reimplementing parts of or whole algorithms in C or Fortran first and then compile, link, and call these from the high level language. In Julia, the programmer can optimize incrementally and immediately see eventual speed improvements within a single language.

\section{Related Work}

There has been a rich history in using JIT compiler techniques to improve the performance of dynamic languages used for technical computing. 
Matlab has had a production JIT compiler since 2002~\cite{matlab2002matlab}.
More recently LuaJIT~\cite{pall2008luajit} and PyPy~\cite{Bolz2009} have shown that sophisticated tracing JIT's can significantly improve the runtime performance of dynamic languages.
Julia's compiler takes advantage of LLVM's JIT for performance, but effort has been directed toward language design and not on improving existing JIT compiler techniques or implementations.
Multimethods, polymorphic types, and multiple dispatch are exemplar language features that allow for greater opportunity for dynamic code optimization.

Multiple dispatch using through multimethods has been explored in a variety of programming languages, either as a built in construct or as a library extension.
A limited sampling of programming languages that support dynamic multiple dispatch are Cecil~\cite{Chambers1992,Chambers1994}, Common Lisp's CLOS~\cite{Bobrow1988}, Dylan~\cite{dylanman}, Clojure~\cite{Hickey2008}, and Fortress~\cite{Allen2011}.
These languages differ in their dispatch rules.  
Cecil, and Fortress employ symmetric multiple dispatch similar to Julia's dispatch semantics.
Common lisp's CLOS and Dylan generic functions rely on asymmetric multiple dispatch, resovling ambiguities in method selection my matching arguments from left to right.
Multimethods not part of the core method system or as a user level library can have user defined method selection semantics.
Such a system is implemented in Clojure~\cite{Hickey2008}, which can reflect on the runtime values of a method's arguments, not just its types, when doing method selection.

Method dispatch in these languages is limited by the expressiveness of the their type systems.
Clojure's multimethods are not a core language construct and only weakly interact with built-in types.
Dispatch in CLOS is class-based and excludes parametric types and cannot dispatch off of Common lisp's primitive value types, limiting its applicability as a mechanism for optimized method selection.  
Cecil's type system supports subtype polymorphism but not type parameters.
Dylan supports CLOS-style class-based dispatch, and also let-polymorphism in limited types, which is a restricted form of parametric polymorphism.
However, Dylan does allows for multiple inheritance. 

Julia is most similar to Fortress in exploring the design space of multiple dispatch with polymorphic types as a mechanism for supporting static analysis. Fortress has additional complexity in its type system, allowing for multiple inheritance, traits, and method selection forcing mechanisms.  This is in contrast to Julia's simpler polymorphic type hierarchy which enforces single inheritance. Static analysis in Fortress is mostly limited to checking method applicability for type correctness.  Julia's use of static analysis is to resolve instances where static dispatch method dispatch is possible and the overhead of full dynamic dipatch can be removed.




\section{Conclusion}
We have describe the design and implementation of Julia's dynamic dispatch mechanism to supporting high-level technical computing programs that also have good performance. In Julia, the combination of dynamic multiple dispatch and on-demand method specialization allows users to write generic code. By providing a uniform language for technical computing programs, Julia provides flexibility and ease of reasoning without requiring the programmer to give up performance.

\bibliographystyle{abbrvnat}
\bibliography{pldi2016,websites}

\end{document}